\documentclass[preprint,12pt]{elsarticle}

\usepackage{amsmath}
\usepackage{amssymb}
\usepackage{color}
\usepackage{makecell}
\usepackage{arydshln}
\usepackage{bbding}
\usepackage{bm}
\usepackage{threeparttable}
\usepackage{tabularx}
\usepackage{multirow}
\usepackage{booktabs}
\usepackage{graphicx}

\journal{Computers in Human Behavior}

\begin{document}

\begin{frontmatter}



\title{Backfire Effect Reveals Early Controversy in Online Media}


\affiliation[1]{organization={Institute of Cyberspace Security},
    addressline={Zhejiang University of Technology}, 
    city={Hangzhou},
    postcode={310023}, 
    country={China}}

\affiliation[2]{organization={Binjiang Institute of Artificial Intelligence},
    addressline={ZJUT}, 
    city={Hangzhou},
    postcode={310056}, 
    country={China}}

\affiliation[3]{organization={Center for Computational Communication Research},
    addressline={Beijing Normal University}, 
    city={Zhuhai},
    postcode={519087}, 
    country={China}}
    
\affiliation[4]{organization={School of Journalism and Communication},
    addressline={Beijing Normal University}, 
    city={Beijing},
    postcode={100875}, 
    country={China}}
    

\author[1,2]{Songtao Peng}

\author[1]{Tao Jin}

\author[1]{Kailun Zhu}

\author[1,2]{Qi Xuan}

\author[3,4]{Yong Min\corref{corresponding}}

\ead{myong@bnu.edu.cn}
\cortext[corresponding]{Corresponding author.}

\begin{abstract}
The rapid development of online media has significantly facilitated the public's information consumption, knowledge acquisition, and opinion exchange. However, it has also led to more violent conflicts in online discussions. Therefore, controversy detection becomes important for computational and social sciences. Previous research on detection methods has primarily focused on larger datasets and more complex computational models but has rarely examined the underlying mechanisms of conflict, particularly the psychological motivations behind them. In this paper, we present evidence that conflicting posts tend to have a high proportion of "ascending gradient of likes", i.e., replies get more likes than comments. Additionally, there is a gradient in the number of replies between the neighboring tiers as well. We develop two new gradient features and demonstrate the common enhancement effect of our features in terms of controversy detection models. Further, multiple evaluation algorithms are used to compare structural, interactive, and textual features with the new features across multiple Chinese and English media. The results show that it is a general case that gradient features are significantly different in terms of controversy and are more important than other features. More thoroughly, we discuss the mechanism by which the ascending gradient emerges, suggesting that the case is related to the "backfire effect" in ideological conflicts that have received recent attention. The features formed by the psychological mechanism also show excellent detection performance in application scenarios where only a few hot information or early information are considered. Our findings can provide a new perspective for online conflict behavior analysis and early detection.
\end{abstract}


\begin{highlights}
\item Combines psychological behavior and conflict detection to develop two new conflict features based on the backfire effect in online media, while the generality of the psychological features is demonstrated on multiple Chinese and English platforms.
\item The general enhancement of our features is shown on several classification algorithms, and the stability and complementarity of the psychological features are explained by quantitative analysis.
\item Psychological features achieve excellent conflict detection performance despite focusing only on hot information (one page) or partial information (early detection).
\end{highlights}


\end{frontmatter}


\section{Introduction}

The definition of controversy or conflict in communication is the interaction of interdependent people who perceive the opposition of goals, aims, and/or values and who see the other party as potentially interfering with the realization of these goals \cite{littlejohn2009}. Debates about religion, politics, intellectual ideology, and even personal preferences occur online and offline daily, and they can disrupt effective communication, leading to cognitive polarization, extremist behavior, and even war \cite{jost2022}. Therefore, ideological conflict has been one of the stresses in modern society that permeates all human activities.

\begin{figure}[ht]
	\centering
		\includegraphics[scale=.45]{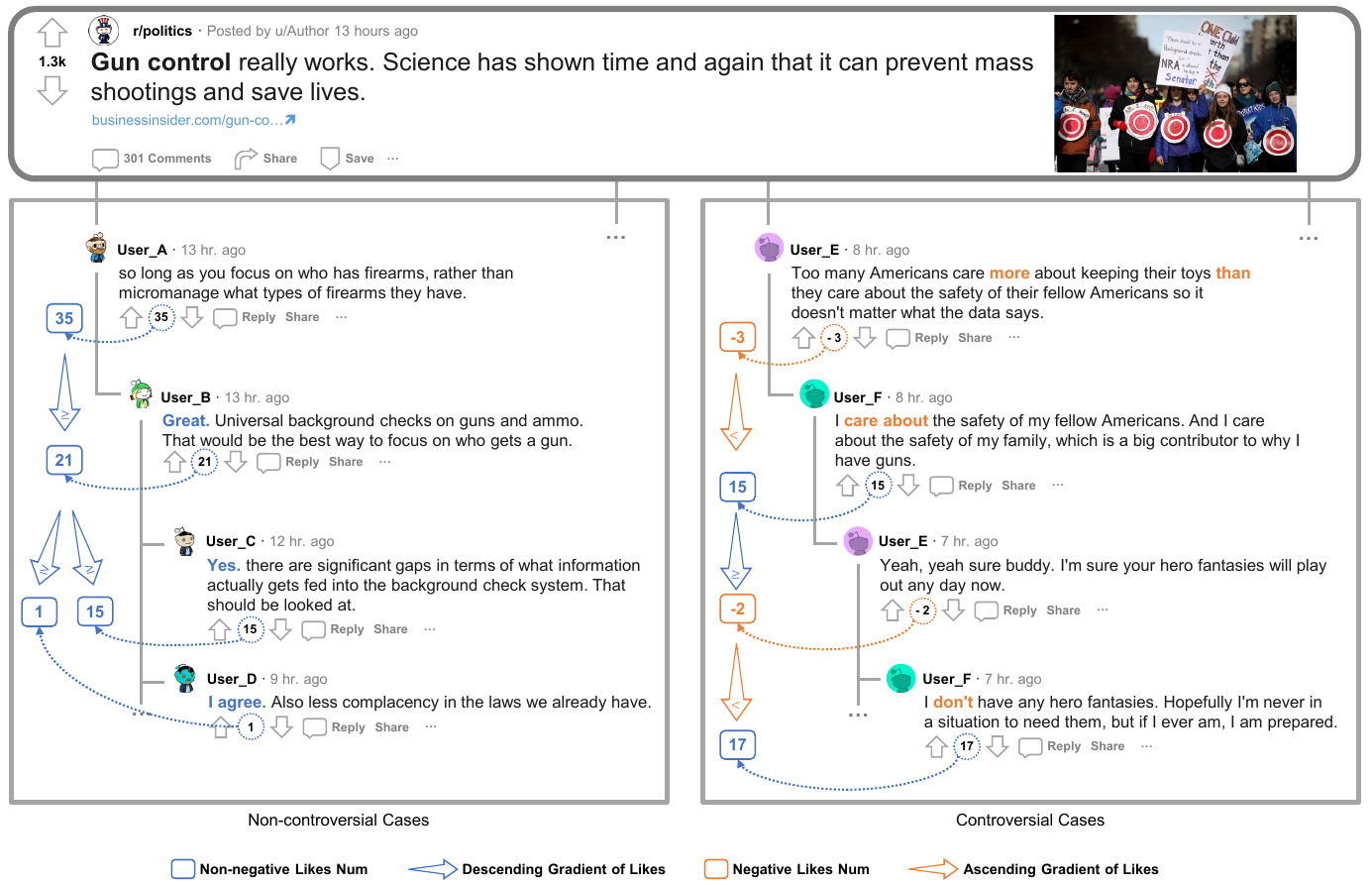}
	\caption{A general illustration of a news or information presentation page in online media. Diverse online media, such as news websites, social media, and forums, can be abstracted into such a page. This page generally includes two parts: the post and related comments. We color-coded features related to conflicting detection on this page and showed the tree structure of comments (orange arrows represent the phenomenon of ascending gradient).}
	\label{fig: post}
\end{figure}

The proliferation of online media has created the potential for more conflicts, and this can be attributed to the nature of the Internet. First, the Internet breaks through traditional geographical restrictions, allowing users with different cultural traditions, religious beliefs, and living habits to contact directly, resulting in unprecedented conflicts \cite{triandis2000}. Second, the Internet makes users feel anonymous, resulting in the lack of necessary personal restraint in the online discussion process. Third, online media, especially social media based on user-generated content, are leaner and lack the necessary regulation, which has led to the proliferation of disinformation and further intensified conflicts \cite{vicario2019}. Finally, artificial intelligence technology in online media could also spark more conflict \cite{stella2018}. These characteristics may result in volatile conversations that are more dynamic in the number of participants and opinions. While this can allow users to engage with each other and discuss differing viewpoints respectfully, it can also lead to escalated disagreements among users.

Given the universality and importance of online conflict, its research has attracted attention from both the social and computational sciences. Research from the social sciences has usually aimed at special controversial issues around critical politics and ideological topics, such as liberalism, equity, and environmental protection \cite{littlejohn2009}. Social science mainly pays attention to the social, psychological, and cultural extension and connotation of these controversial issues and how users participate in discussing these issues \cite{de2012}. In contrast, computational science has primarily focused on creating detection models in online media \cite{dori2015}. Facing the vast content of online media, efficient and accurate conflict detection methods are the technical basis for carrying out related research and the premise for effective conflict management.

Current conflict detection models are trained using features such as natural language, social networks, and user interactions to identify conflicting content at different levels, including topic-level, post-level, and interaction-level \cite{al2018}. However, the existing research still has certain limitations. With the development of artificial intelligence technology, natural language processing (NLP) methods have become the most natural choice for conflict detection. Due to natural languages' inherent complexity and uncertainty, these models depend highly on language types and the pertinence of corpus \cite{bide2021}. As a result, these models cannot be used directly in other languages or topics. Unlike language-based models, detection models based on social networks and user interaction have better generalizability \cite{coletto2017, saveski2021}. However, they usually exhibit severe data dependence, such as large-scale friend relationships or complete discussion processes. The dilemma between universality, usability, and effectiveness requires further development of existing detection models, especially to discover more effective representations of online conflicts.

We aim to resolve this dilemma at the post-level (whether a piece of news or a post will create conflict) by finding new conflicting features (Fig. \ref{fig: post}). Instead of analyzing and detecting a post using all the information of the post, we use the data that can be obtained within the "one page" containing the content and related comments of the post to avoid relying on massive data. At the same time, we use the post and the comment data with the earlier posting time to achieve early conflict detection. More thoughtfully, we consider the diversity of online media (including forums, and news websites) and only select commonly available data for the various platforms. By considering the psychological mechanisms behind features, we have introduced two novel features for conflict detection across media platforms from different languages and types. We quantify the value of the new features by comparing the discernibility with other frequently used features. The new features can facilitate the development of better detection models, and this feature-mining method based on psychology and behavior also provides new pathways for future social media research. The main contributions of this paper include: 
\begin{enumerate}[(1)]
\item Combines psychological behavior and conflict detection to develop two new conflict features based on the backfire effect in online media, while the generality of the psychological features (backfire effect) is demonstrated on multiple Chinese and English platforms.
\item The general enhancement of our features is shown on several classification algorithms, and the stability and complementarity of the two psychological features are explained by quantitative analysis.
\item Psychological features are suitable for more efficient application scenarios, achieving excellent conflict detection performance despite focusing only on hot information (one page) or partial information (early detection).
\end{enumerate}

The remainder is organized as follows. Section \ref{sec: related_work} provides an overview of the related work on conflict detection, focusing on different categories of features. Section \ref{sec: methodology} outlines multilingual and multi-platform datasets, candidate features, and the detection method. Section \ref{sec: result} presents the analysis of the experiments. Finally, Section \ref{sec: conclusion} summarizes the key findings and highlights future research directions.

\section{Related Work}
\label{sec: related_work}
Many features have been used for conflict detection, which can be divided into four categories: user characteristics, text-based features, social networks, and discussion interactions. In this section, we briefly summarize the research on these features.

\subsection{User Characteristics}
Reviewing previous research, distinct population characteristics can lead to different types of participation in conflict. For example, Eagly and Steffen showed that there was a big difference between male and female when it came to aggressive behavior \cite{eagly1986}, and Holt and DeVore found that men were more likely to report using forcing than women in individualistic cultures and with regard to organizational role, men were more likely than women to choose a forcing style with their superiors \cite{holt2005}. Triandis demonstrated that the differences between culture and race had profound implications for the probability of conflict and the type of conflict \cite{triandis2000}. Cheng et al. showed that antisocial behavior in online groups is a relatively stable individual difference \cite{cheng2015}. Levy et al. also proved that user activity is also closely related to conflict behaviors \cite{levy2022}. In addition, the previous behavior of the users participating in the discussion can also be used to detect a future conflict. Saveski et al. found that the structural characteristics of the conversation were also predictive of whether the next reply posted by a specific user will be toxic or not \cite{saveski2021}. Their findings can be used to detect early signs of toxicity and potentially steer conversations in a less toxic direction. However, the acquisition of user characteristics has certain limitations. Often, user information is not provided on the same screen as posts or comments and needs to be collected outside. At the same time, with the increasingly strict privacy policy of the platform, the moral hazard of obtaining user characteristics is also increasing.

\subsection{Text-based Features}
In conflict detection, the text data of posts or comments are usually easy to obtain, which is conducive to fully mining sentiments, emotions, and opinions \cite{al2018, kumar2018, choi2020}. These papers have analyzed the intensity of emotions and sentiments, showing that these text-based features are strong indicators of controversy. Weingart et al. analyzed conflict spirals in the workplace which were influenced by conflict expressions varying in directness and oppositional intensity \cite{weingart2015}. Zhang et al. predicted from the very start of a conversation whether it would get out of hand within Wikipedia talk page discussions through various text-based features such as politeness strategies and prompt types \cite{zhang2018}. However, the text-based features have certain limitations, including the type of language and dependencies of prior knowledge and corpus. In addition, The scene of short texts in social media also increases the difficulty of emotion and semantic recognition, e.g., "Ha Ha" may indicate happiness or ridicule simultaneously. In sum, the conflict has been connected to increased intensity and negative emotions represented by text-based features. 

\subsection{Social Networks}
In addition to user characteristics, researchers study the conflict commenter's social network structure. Conover et al. demonstrate the existence of controversial structures in Twitter through network modularity and graph partitioning based on user relationships and behavioral relationships in social networks \cite{conover2011}. Morales et al. quantified the controversial nature of topics through opinion leaders in the network structure \cite{garimella2017}. Akoglu et al. constructed a bipartite graph based on user relationships to quantify the contentiousness of information \cite{akoglu2014}. Coletto et al. studied users' connections in the context of controversial threads \cite{coletto2017}. To do so, they analyzed local network patterns of user-follower and user-reply graphs. Their findings showed that controversial interactions are less likely between users who follow each other on social media. de Dreu argued that conflicts occurred more often in intergroups than in interpersonal \cite{de2010}. Many of these conflicts are identity-based, in which people believe that the group or subgroup they identify with (e.g., ethnicity, race, religion, or political party) is superior to an outgroup \cite{mckeown2009, tajfel1974, triandis2000}. While many conflicts originate from differences between intergroups, these intergroups conflicts can evolve into interpersonal ones \cite{labianca1998}, and frequent pleasant interactions can reduce intergroup conflicts \cite{pettigrew2011}. 

Besides, some scholars combine network structure with text semantics. Zhong et al. proposed a controversial microblog detection algorithm based on Graph Convolutional Network \cite{zhong2020}. This algorithm builds a topic-microblog-comment network based on behavioral relationships and combines the textual information of topics and comments to detect controversial details. Benslimane et al. proposed a method for contentious information detection based on graph structure and textual features \cite{benslimane2021}. The technique embeds graph representations (including text features) into feature vectors \cite{xuan2019subgraph}, considers the attention mechanism when calculating user nodes, fuses the importance of neighbor nodes, and finally performs the classification task through the Graph Neural Network.

Although social networks reveal the connections behind comment conflicts, acquiring and analyzing large-scale social network data is relatively inefficient. Moreover, not all online media are based on social networks; for example, news sites often do not offer friendship among users.

\subsection{Discussion Interactions}
In addition to social relations, users will also discuss and interact through comments below the post. The structure of this interaction has also been shown to be related to the conflict. Coletto et al. not only analyzed the relationship between user-follower and user-reply networks but also investigated intensity through a different perspective by assuming controversial topics may generate "dense" discussions in time so that the inter-reply rate for these conversations is lower (i.e., more rapid replies) than those of non-controversial ones \cite{coletto2017}. Discussion interaction contains the psychological and behavioral characteristics of users expressing opinions under conflicting issues, and its potential needs further exploration.

\section{Methodology}
\label{sec: methodology}

\subsection{Data Processing}
To evaluate the discriminative power of features in different languages and types of online media, we collected three datasets from various sources: Reddit (English), Toutiao (Chinese), and Sina (Chinese). Each dataset contains posts from multiple sets of conflicting and non-conflicting topics from their respective platforms.

The English-based media platform is Reddit. For the Reddit platform, we use the May 2015 public dataset from the kaggle.com platform. We filtered controversial topics from two subreddits using keywords according to the list from the library of Shippensburg University. The two topics mentioned above, Gun and War, and the content under the "DebateReligion" subreddit, make up the three controversial topics. Meanwhile, we get three mild topics of Shopping, Scenery, and Music around the three subreddits "Random\_Acts\_Of\_Amazon", "Watches", and "Guitar". The dataset contains the complete information of posts, comment information, and the relationship between each comment. To ensure that the posts under each topic have a complete relationship structure and diverse information attributes, we mainly focus on posts with more than 50 comments. Finally, the dataset contains 542 news and 110,176 comments (The main data in Table \ref{tab: dataset}).

The Chinese-based media platforms are Toutiao and Sina. Toutiao or Jinri Toutiao is a news and information platform based on user-generated content (UGC) and recommendation algorithms. The dataset includes news information, comment information, and the relationship of comments. The period of the data is from January 2019 to December 2019. In Toutiao, we choose the topics of Huawei, NBA, and LOL that have apparent controversy, and the topics of life, food, and travel with minor controversy. The raw data of this dataset is collected by the web crawler, and then we clean the raw data to remove the empty and duplicate values. In the end, there were 647 news and 119,725 comments in the dataset. Sina is one of the most influential news platforms on the Chinese Internet. The period of crawled data is from January 2019 to December 2019. We select the same controversial topics as the Toutiao, as well as mildly controversial topics such as science, music, and movies, for a total of 715 news items and 81,337 comments (The test data in Table \ref{tab: dataset}).

\begin{table}[ht]
\caption{All datasets. Three Chinese and English media platforms (Reddit, Toutiao, Sina) are included, and six topics are extracted from each platform. Each topic contains the attributes of controversy, the total number of posts, and comments. Reddit is the main experimental data. Toutiao and Sina are used for experimental testing. }
\label{tab: dataset}
\renewcommand\arraystretch{1.2}
\resizebox{1.0\linewidth}{!}{
\begin{tabular}{cccccc}
\hline
Type & Media & Topic & Controversy & Post & Comment  
\\ \hline
\multirow{6}{*}{\begin{tabular}[c]{@{}c@{}}Main\\ Data\end{tabular}} & \multirow{6}{*}{\begin{tabular}[c]{@{}c@{}}Reddit\\ (English)\end{tabular}} 
& Gun  &  \Checkmark   & 79  & 19,371  \\
&   & War  &   \Checkmark      & 66  & 17,556      \\
&     & Religion   &    \Checkmark     & 122  & 25,663         \\
&    & Shopping        &    \XSolidBrush    & 111  & 26,503      \\
&     & Scenery    &   \XSolidBrush       & 71  & 7,805        \\
&      & Music       &   \XSolidBrush       & 93  & 13,278         \\ \hline
&       & \multicolumn{2}{c}{Toutiao(Chinese)} & \multicolumn{2}{c}{Sina(Chinese)} \\
&      & Topic           & Post/Comment       & Topic        & Post/Comment       \\ 
\cline{3-6} 
\multirow{6}{*}{\begin{tabular}[c]{@{}c@{}}Test\\ Data\end{tabular}} & 
\multirow{3}{*}{\begin{tabular}[c]{@{}c@{}}Controversy\\ \Checkmark \end{tabular}}                                       
& Huawei     &  111/25,317      &   Huawei    &    105/16,132        \\
&         & NBA       &  104/22,695        & NBA    &   145/15,702    \\
&        & Football        &  96/21,747     &   Football    &   171/18,298      \\ 
\cline{3-6} 
& \multirow{3}{*}{\begin{tabular}[c]{@{}c@{}}Controversy\\ \XSolidBrush \end{tabular}}   
& Life     &  104/13,434    &   Science     & 84/7,910       \\
&         & Food   &   115/26,518     &    Music    & 78/8,914     \\
&         & Travel          &    117/10,014       &  Movie      &  132/14,381  \\ \hline
\end{tabular}
}
\end{table}

Our dataset has a clear hierarchical structure. \textbf{Topics} (subreddits) allow users to come together to discuss topics of interest, share relevant content and resources, interact with others, and build a community focused on a specific topic. \textbf{Posts} are the body for our subsequent analysis, containing information such as the post topic, the post time, and the user ID. \textbf{Comments} are the key unit of our analysis and contain information such as comment text, comment time, and number of likes.

This paper uses $P$ to denote a post, that is:
\begin{equation}
    P=\left\{\text { Topic }, P_{i d}, P_{t}, T\right\}
\end{equation}
Where $P_{id}$ is the individual ID of a post, $P_t$ is the time when a post was published, and $T$ is the post content.

We define a comment as $C$, that is:
\begin{equation}
    C=\left\{P_{i d}, C_{i d}, C_{t}, C_{l}, T, C_{p i d}\right\}
\end{equation}
Where $P_{id}$ is the ID of a post to which the comment belongs, $C_{id}$ is the comment ID, $C_t$ is the time that the comment was posted, $C_l$ is the number of likes for a comment, $T$ is the textual content, and $C_{pid}$ is the ID of the parent text of the comment (i.e., the previous level of text to which the current comment belongs).

\begin{figure*}[ht]
    \includegraphics[width=\textwidth]{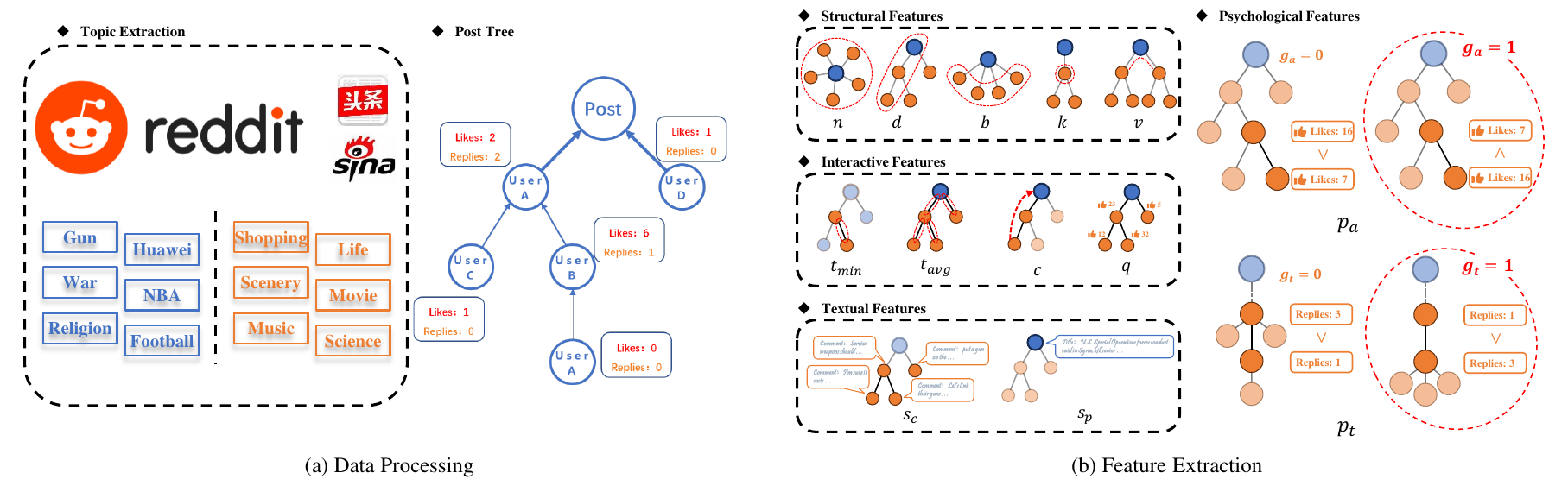}
    \caption{The two key parts of the methodology. (a) Data processing realizes topic extraction and post-tree construction. (b) Feature extraction describes four classes totaling 13 features, including our newly designed psychological features.}
  \label{fig: model}
\end{figure*}

The form of a graph can be good for representing and understanding posts. We define a node set $V={v_0,v_1,v_2,\cdots,v_n}$, where $v_0$ represents the post $P$, $v_i(i\neq0)$ denotes the comments, and $n$ is the number of comments on the post. Then, we define the edge set $E={e_1,e_2,\cdots,e_n}$. Where, if the parent text ID $C_{pid}$ of node $v_j$ points to node $v_i$, then the connected edge $e_j=<v_i,v_j>$ is formed. From this, we can construct a comment graph $G=(V, E)$ for different data in terms of posts, and Fig. \ref{fig: model} (a) is a simple schematic diagram.

\subsection{Feature Extraction}
Based on multilingual and multi-platform datasets, we choose the common features among them as comparison objects. This web page usually contains the post title, content, and comments on the post and their replies. Therefore, we consider structural, interactive, and textual features. Note that we did not consider user characteristics, as this was far beyond the data presented by the information presentation page. We aim to identify conflicting posts with limited information quickly.

First, we consider features related to comment structure. A post and its comments can form a tree structure, where the root is the post and other nodes are comments, and these nodes are connected by comments or reply relationships. Based on the tree structure, we can extract the following features \cite{vosoughi2018}:
\begin{itemize}
  \item \textbf{Size ($n$)}. The size of a tree corresponds to the number of nodes (without root) in that tree. For example, the size of the demo post in Fig.\ref{fig: model}(a) is 5.
  
  \item \textbf{Depth ($d$)}. The depth of a node is the number of links from the node to the root. The depth of a tree is the maximum depth of the node in all nodes. In other words, the depth of a tree, $d$, with $n$ nodes is defined as: 
  \begin{equation}
    d=max(d_i), 0\leq i\leq n
  \end{equation}  
  where $d_i$ denotes the depth of node $i$. For example, in the Fig.\ref{fig: model}(a), $d=3$.
  
  \item \textbf{Breadth ($b$)}. The breadth of a cascade is a function of its depth. At each depth, the breadth is the number of nodes at that depth. The breadth of a tree is its maximum breadth at all depths. For a tree with depth $d$, the breadth, $b$, is defined as:
  \begin{equation}
    b=max(b_i), 0\leq i\leq d
  \end{equation}  
  where $b_i$ denotes the breadth at depth $i$. For example, in the Fig.\ref{fig: model}(a), $b=2$.
  
  \item \textbf{Average degree ($k$)}. In a tree, the degree of a node is only the number of direct children of a node. Therefore, the average degree of a tree is defined as:
  \begin{equation}
    k=\frac{\sum_{i=1}^{n}k_i}{n}, 0\leq i\leq n
  \end{equation}  
  where $k_i$ denotes the degree of node $i$. For example, in the Fig.\ref{fig: model}(a), $k=1.2$.

  \item \textbf{Structural Virality ($v$)}. The structural virality of a cascade, as defined by Goel et al. (2015), is the average distance between all pairs of nodes in a cascade. For a cascade with $n>1$ nodes, the virality $v$ is defined as: , 
  \begin{equation}
    v=\frac{1}{n(n-1)}\sum_{i=1}^{n}\sum_{j=1}^{n}d_{ij}
  \end{equation}
  where $d_{ij}$ denotes the length of the shortest path between nodes $i$ and $j$. For example, in the Fig.\ref{fig: model}(a), $v=1.1$.
\end{itemize}

Second, we considered text-based features, including:
\begin{itemize}
  \item \textbf{Intensity of comment emotion ($s_c$)}.  We combine two sentiment methods, Vader and SnowNLP, to achieve efficient semantic analysis of Chinese and English \cite{min2015, chou2020}. Based on the sentiment dictionary, a sentiment score $c_i$ can be calculated for each comment (node $i$), and then the comment emotional intensity of a post can be obtained by the following equation:
  \begin{equation}
    s_c=\frac{\sum_{i=1}^{n}c_i}{n}, 0\leq i\leq n
  \end{equation} 
  \item \textbf{Intensity of post emotion ($s_p$)}. The emotional intensity of a post is also based on the above sentiment dictionary, and the sentiment score of an entire post content is calculated as $s_p$.
\end{itemize}

Third, the features of discussion interaction are also considered \cite{coletto2017}:
\begin{itemize}
  \item \textbf{Reply time ($t$)}. For each reply link $e_x=\left\langle i,j \right\rangle$ in a post, the time elapsed between the parent comment (node $i$) and its child (node $j$) is reply time ($t_ij$). We consider average ($t_{avg}$) and minimum ($t_{min}$) value of all $e_x$.
  \item \textbf{Comment density ($c$)}. For a post, the time interval from its publication to the latest comment is $\delta$, then the density of comments can be defined as:
  \begin{equation}
    c=\frac{n}{\delta}
  \end{equation}  
  where $n$ denotes the number of comments.
\end{itemize}

Unlike the previous features, we consider the number of likes of comments \cite{levy2022}. We first define $L_i$ to denote the number of likes received by comment $i$ and consider the average number of likes received by all comments:
\begin{equation}
  q=\frac{\sum_{i=1}^n L_i}{n}
\end{equation}

Second, We hypothesize that when users view a comment with an opposing opinion, they are more willing to press "like" to reply to criticizing the comment. So, we define the gradient function for each reply link $e_x=\left\langle i,j \right\rangle$ as:
\begin{equation}
  g_{a}(e_{x})=\begin{cases}
    1,&\text{if } L_i-L_j<0;\\
    0,&\text{otherwise.}
  \end{cases}
\end{equation}
For all links, we can obtain the proportion of ascending gradients as:
\begin{equation}
  p_a=\frac{\sum_{x=1}^m g_{a}(e_{x})}{m}
\end{equation}
where $m$ is the number of reply links in the post.

Furthermore, we expand the concept of ascending gradient to encompass the difference in the number of replies across comments, which we refer to as a "tier ascending gradient". First define $R_i$ for each comment, i.e., the total number of comments that directly reply to comment $i$. Then similar to Equations 8 and 9, we can obtain the functional representation of the tier ascending gradient as:
\begin{equation}
  g_{t}(e_{x})=\begin{cases}
    1,&\text{if } R_i-R_j<0;\\
    0,&\text{otherwise.}
  \end{cases}
\end{equation}
\begin{equation}
  p_t=\frac{\sum_{x=1}^m g_{t}(e_{x})}{m}
\end{equation}

All features are also summarized and described in Fig. \ref{fig: model} (b). For convenience, we subsequently abbreviate the ascending gradient feature as AG and the tier ascending gradient feature as TAG and refer to them uniformly as psychological features (this will be analyzed in the experimental section to explain the psychological property of our proposed features). Finally, all the features and their descriptions that we have compiled are shown in Table \ref{tab: features}.

\subsection{Model Construction}
LightGBM is a gradient enhancement framework based on the decision tree algorithm proposed by Microsoft Research in 2017 \cite{ke2017lightgbm}, which is widely used in classification or regression tasks. LightGBM optimizes the defects of Gradient Boosting Decision Tree (GBDT) that exist in high time and space consumption, supports efficient parallel training, and has the advantages of faster training speed, lower memory consumption, and a wider amount of processed data. It can also be called the GBDT algorithm using Gradient-based One-Side Sampling (GOSS) and Exclusive Feature Bundling (EFB), so first is a brief introduction to the core strategies GOSS and EFB.

\textbf{GOSS}: The GOSS algorithm starts from the point of view of reducing samples, excludes most of the samples with small gradients and uses only the remaining samples to calculate the information gain, realizing the reduction of the data volume while guaranteeing the balance of accuracy. Therefore, GOSS only retains the data with larger gradients when sampling data, but if all the data with smaller gradients are directly discarded, it will definitely affect the overall distribution of the data. Therefore, GOSS firstly arranges all the values of the features to be split in descending order of absolute value size and selects $a*100\%$ of the data with the largest absolute value. Then $b*100\%$ of the remaining smaller gradient data are randomly selected. This $b*100\%$ of the data is then multiplied by a constant $(1-a)/b$ so that the algorithm focuses more on the under-trained samples and doesn't change the distribution of the original dataset too much. Finally, this $(a+b)*100\%$ of the data is used to calculate the information gain.

\textbf{EFB}: In order to solve the sparsity of high-dimensional data, mutually exclusive features are bundled to achieve feature dimensionality reduction without losing information. The process is mainly divided into four steps:
\begin{enumerate}[1)]
\item Take features as the node of a graph, for the features that are not mutually exclusive are connected (i.e., there exist samples that are not 0 at the same time), and the number of samples whose features are not 0 at the same time are used as the weights of the edges.
\item Sort the features in descending order based on the degree of the nodes, with a larger degree indicating that the features are in greater conflict with other features (the less it will be bundled with other features).
\item Set the maximum conflict threshold $K$, traverse the existing feature clusters and if it is found that the number of conflicts for the feature to be added to the feature cluster will not exceed the maximum threshold $K$, then the feature is added to the cluster. Otherwise, create a new feature cluster and add the feature to the newly created cluster.
\item Separate the original feature from the merged feature by adding an offset constant.
\end{enumerate}

Among them, the first three steps solve the problem of determining which features should be bundled together, and the last step solves the problem of how to bundle (merge) multiple features into one.

In addition to this, the Histogram-based decision tree strategy allows the algorithm to run with a smaller memory footprint and less computational cost. The Leaf-wise strategy with depth constraints ensures the high efficiency of the algorithm while preventing overfitting. So, LightGBM is our main algorithmic model.

\section{Results and discussion}
\label{sec: result}

In this section, detailed experiments are conducted around the task realization designed in Section \ref{sec: methodology}, and the results are presented in three parts as follows: algorithm evaluation results, importance evaluation results, and application evaluation results.

\subsection{Controversy detection results}
If suitable features are chosen, even simple classification models can achieve excellent detection results. Based on this viewpoint, we first validate the enhancement effect of psychological features on different controversial topic detection algorithms, which include interval-based SVM, distance-based KNN, probability-based LR, tree-structure-based DT, integrated learning-based GBDT and LightGBM, the results of which are shown in Table \ref{tab: algorithm}. The best testing metrics are achieved by all classical classification algorithms when using full features, reaching 93.25\% in SVM and LightGBM. When the psychological features are removed, all models have a significant decrease in detection accuracy, with a maximum difference of up to 10\%. Then, it also has some performance decrease after removing the text features. Meanwhile, we conducted experiments on the toutiao and sina datasets using the current best-performing LightGBM algorithm, and the results were optimal with the addition of psychological features. In the English platform, text-based features also show a better boost. Conflict detection intuitively is supposed to be the interaction of opposite emotions between utterances, so it is reasonable that the importance of text-based features is significant while the maturity of existing NLP technologies in English. However, the characteristic of text-based features is that with the increase of sample size, its effect will be more obvious, but at the same time, it is also a limitation that it is too dependent on the data sample size and emotion recognition technology, which fluctuates a lot. In the Chinese platform, the text features do not show obvious effects (this is partly attributed to the fact that there are still some limitations in the sentiment analysis technology of Chinese short texts). Unlike the volatility accompanying structural, interactive, and textual features, our psychological features have a general enhancement effect on different algorithms and platforms, which is worthy of subsequent in-depth exploration and research.

\begin{table}[ht]
\caption{Algorithm evaluation results. The detection performance of different features in different classification models is evaluated on the Reddit dataset. The detection performance of different features in the LightGBM model is verified on the Toutiao and Sina datasets.}
\label{tab: algorithm}

\resizebox{1.0\linewidth}{!}{
\begin{tabular}{cccccc}
\toprule
\multirow{2}{*}{Platform} & \multirow{2}{*}{Method} & \multicolumn{4}{l}{Feature}                 \\
                          &                         & Structure & Interaction & Text & Psychology \\ \hline
\multirow{6}{*}{Reddit}   & SVM                     &  80.98    &  80.37      &  80.98    &  \textbf{93.25}          \\
                          & KNN                     &  82.82    &    82.82    &   85.28  &   \textbf{89.57}            \\
                          & LR                      &  80.37    &    79.75    &   80.37  &   \textbf{91.41}            \\
                          & DT                      &  74.23    &    82.82    &   76.68  &   \textbf{85.27}            \\
                          & GBDT                    &  82.82    &    84.66    &   86.50  &   \textbf{90.80}            \\
                          & LightGBM                &  84.05    &    84.05    &   85.28  &   \underline{\textbf{93.25}}           \\ \cline{2-6} 
Toutiao                   & LightGBM                &   77.84   &   89.18     &   88.66  &  \underline{\textbf{90.21}}           \\
Sina                      & LightGBM                &   78.22   &   82.18     &   82.67  &  \underline{\textbf{83.66}}           \\ 
\bottomrule
\end{tabular}
}
\end{table}

\begin{figure*}[ht]
  \centering
  \includegraphics[width=\textwidth]{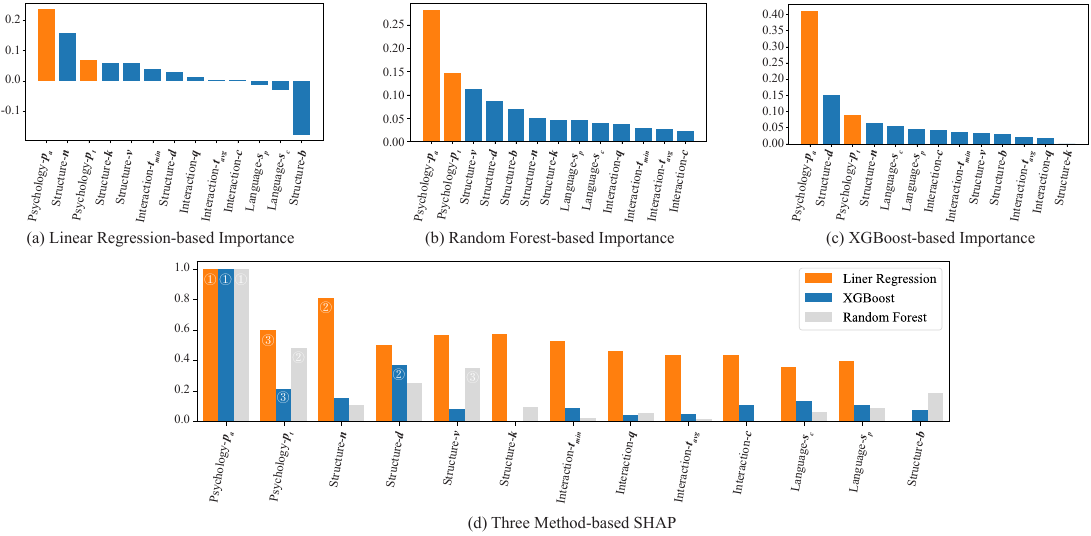}
  \caption{The comparison of importance of size ($\bm{n}$), depth ($\bm{d}$), breadth ($\bm{b}$), average degree ($\bm{k}$), structural virality ($\bm{v}$), comment emotion ($\bm{s_c}$), post emotion ($\bm{s_p}$), minimum reply time ($\bm{t_{min}}$), average reply time ($\bm{t_{avg}}$), comment density ($\bm{c}$), average ups ($\bm{q}$), ascending gradients ($\bm{p_a}$), and tier ascending gradients ($\bm{p_t}$) in Reddit dataset.}
  \label{fig: reddit}
\end{figure*}
\subsection{Importance evaluation results}
\subsubsection{Quantitative analysis}
At the post level, we mixed all posts regardless of topic and assessed the importance of features using three classification algorithms. The results show that in all algorithms and platforms, the two features ($p_a$, $p_t$) based on the ascending gradient are far more critical than the others in distinguishing controversies, and are generally in the top three positions of importance (Fig. \ref{fig: reddit}(a)$\sim$(c)). In the RF algorithm, the two features are in the first and second positions respectively, with values up to 0.27 and 0.14. In the XGBoost algorithm, the gap between the first feature and the second feature can even reach nearly three times. The SHAP \cite{lundberg2017} values of each feature also show the same effect (Fig. \ref{fig: reddit}(d)), i.e., $p_a$ is generally in the first position, and psychology-based features commonly in the top three positions. Overall, for the discriminative ability of conflicting posts, the value of ascending gradient features is significantly better than other features, while $p_a$ is generally better than $p_t$.

Second, we used the Kolmogorov-Smirnov test to compare the differences in the distribution of each feature across the three conflicting and non-conflicting topics of the Reddit dataset (Appendix Fig. \ref{fig: feature_compare}). The results showed that among all five comment structure features, only breadth and structural virality differed significantly between conflicting and non-conflicting topics ($p<0.05$). Also, there were significant differences between specific topics in terms of size and depth. For text-based features, the sentiment intensity of comments mostly showed certain differences when comparing conflicting and non-conflicting topics, which is consistent with previous findings \cite{choi2020}. We found a significant difference in the distribution of average replay time and average ups for discussions between conflicting and non-conflicting topics, with slightly shorter phrases for controversial topics and non-controversial topics.

\begin{figure*}[ht]
  \label{fig:reddit}
  \centering
  \includegraphics[width=0.9 \textwidth]{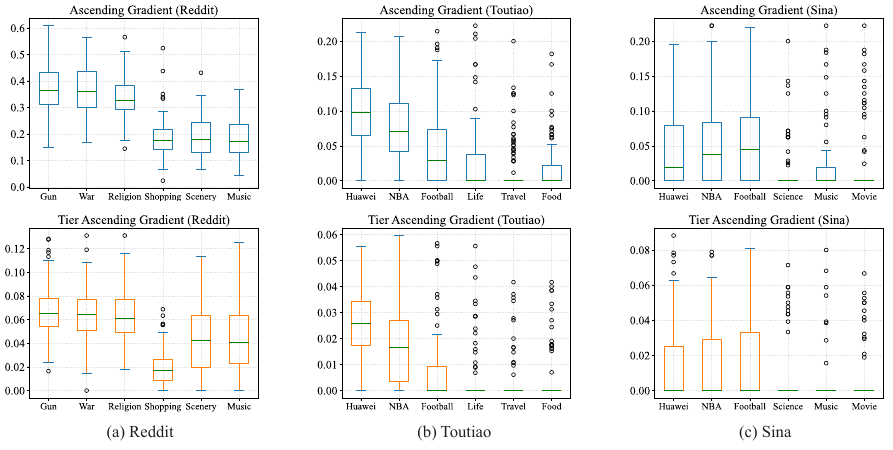}
  \caption{Comparing the distribution of ascending gradients ($p_a$) and tier ascending gradients ($p_t$) of different topics in three media platforms.}
  \label{fig: AG}
\end{figure*}

Among the different types of topics in the Reddit platform, for the gradient-related features (AG features in blue and TAG features in orange in Fig. \ref{fig: AG}), there is a clear distributional difference, i.e., the feature values of non-controversial topics are generally lower than those of controversial features. In particular, the average value of the AG feature in controversial topics is about 20\% higher than that of non-controversial topics, which shows that the feature is a common situation among different types of topics. To verify that the difference in the average number of likes and replies to comments is also a common phenomenon across different platforms, we conducted a validation on two news platforms, Toutiao and Sina, whose results show a very obvious difference as well. Meanwhile, because the news platforms are mostly dominated by first-level comments, the feature magnitude of the controversial topics is a few times different from that of Reddit platforms, and the gradient feature values in terms of non-controversial topics are dominated by 0. In summary, the controversial topic gradient feature values are generally higher than the non-controversial topics, with significant fluctuations.

To summarize, ascending gradient-based features are significant and valuable in detecting controversies, and they can be applied to various media platforms in both Chinese and English.

\subsubsection{Psychological interpretation}
Why does ascending gradient perform well in discriminating between conflicting topics and posts? From a psychological point of view, the backfire effect could explain the emergence of the ascending gradient of likes and replies. A recent wave of studies suggested exposure to those with opposing political views may create backfire effects that stimulate the user's cognitive resistance and exacerbate political polarization \cite{nyhan2010, bail2018}. In theory, the deeper the level of comments, the more difficult it is to be exposed and the lower the likelihood of getting likes or replies. But in controversial posts, when some users (who may be silent by default) first see a comment that challenges their opinion at a lower level, and then, due to the backfire effect, they will be more insistent on their opinions. They will be motivated to seek psychological compensation. Therefore, when these users find that other users are refuting or fighting against this comment, they are more likely to be encouraged to express their desire to express it through likes or replies. In the upper level, comments that are consistent with the user's point of view are not easy to get likes or replies from the type of users because they are not stimulated by the backfire effect. Since this excitation is consistent with user groups with opposing opinions, it leads to the emergence of ascending gradient situations in controversial posts. Due to this, we refer to the proposed gradient features as psychological features.

The backfire effect reminds us that in the discussion of controversial topics, users will be repeatedly exposed to opposite opinions, which will inevitably change their psychological state. It can be considered that the increase in emotional intensity and negativity in the discussion is related to this, but emotion recognition relies on natural language processing technology, which has certain limitations. In a word, the user's mental state and behavior leading to conflict is a causal relationship, while identifying conflict based on techniques such as natural language processing is a process of data analysis. Next, we apply psychological features to several real-life application scenarios.

\subsection{Application evaluation results}
In view of the above sufficient validation, we focus on more efficient application scenarios applied to topic controversy detection for a small amount of hot information (One page) or partial information (Early detection). We demonstrate that our approach achieves efficient conflict detection with minimal information while providing an interpretable analysis of its performance.

\subsubsection{One page}

\begin{figure}[ht]
  \centering
  \includegraphics[width=\textwidth]{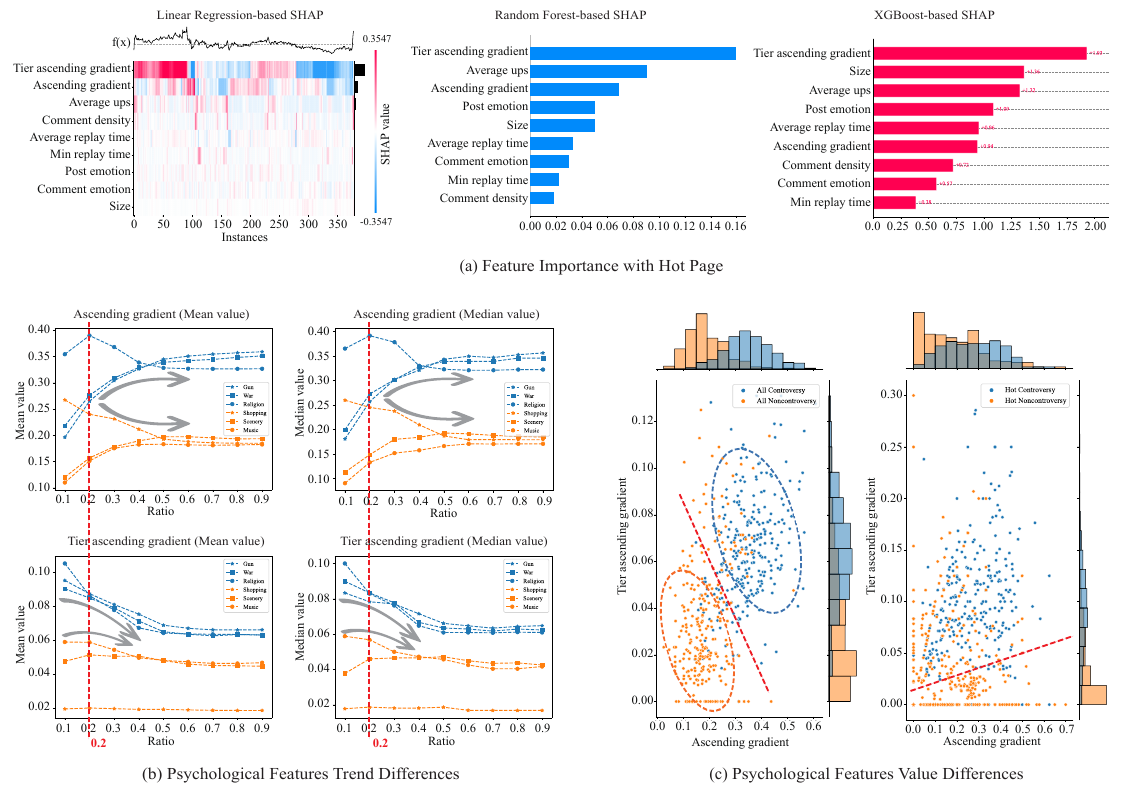}
  \caption{Analysis of feature importance and interpretation based on hot comments. (a) Three algorithms are used to evaluate the importance of 9 features under hot comments. (b) The trend of the Ascending gradient and Tier ascending gradient feature values at different scale reviews is assessed using the mean and median, respectively. (c) Evaluate the differences in the distribution of Ascending gradient and Tier ascending gradient feature values under hot comments (ratio=0.2) and all comments (ratio=1.0), respectively.}
  \label{fig: hot}
\end{figure}

For each post page under the social platform, the first thing that is displayed is the highly liked comments to get an idea of the hot topics being discussed at the moment. As a result, most of the debates are also centered around the hot comments, while the hidden comments due to low attention have less influence on the judgment of topic conflicts. Therefore, we propose "one-page", i.e., using a small amount of intuitively visible hot information to achieve highly accurate conflict detection. For the experimental validation of hot comments, we only focus on comments with a high number of likes (top 20\%) and their reply pairs in the current post. Meanwhile, the four structural features are not used because the global tree structure is lost. With only 20\% of the hot comment information used, the LightGBM algorithm still has 87.12\% detection accuracy with 9 features. We again used the above three methods to assess the importance of features and the results are shown in Fig. \ref{fig: hot}(a). Psychological features are generally in the top position and their values are significantly better than other features.

It is worth noting that from the results we find that TAG features are at the most important position in all the evaluations, which is different from the trend in Fig. \ref{fig: reddit}. So for the case that TAG features are significantly better than AG features, we did a more in-depth experimental validation. Fig. \ref{fig: hot}(b) represents the trend of the same feature under different ways of processing the feature values (vertical, taking the average or median of the feature values of all articles on the same topic). It can be clearly seen that as the ratio increases, the variability between samples of different classes under the AG feature has an expanding trend, while the TAG feature is a contracting trend. This intuitively proves the validity of AG features and the superiority of TAG features when the ratio is 0.2. The reason for the general superiority of AG features when evaluating the importance of features in Fig. \ref{fig: reddit} (ratio = 1.0) is also confirmed by Fig. \ref{fig: hot}(b). We also explain the analysis in terms of the distribution of feature values (Fig. \ref{fig: hot}(c) shows the case of posts with complete information on the left, and the case of posts with hot information on the right), where the horizontal and vertical coordinates of the figure represent the two psychological features, and each point represents a post. It can be clearly seen that when the information is complete, the boundary between the two types of posts is clear and the AG feature brings more distinctive differentiation. When the information is incomplete, a large number of non-controversial posts (in orange), are distributed all over the bottom, which makes the vertical TAG feature work more clearly than the horizontal AG feature. Overall, both AG features and TAG features can play an important role in controversy detection, while also complementing each other in different situations.

\subsubsection{Early detection}


\begin{table}[ht]
\centering
\caption{Evaluate the conflict detection performance in different feature modes using comment information over different time periods.}
\label{tab: application}

\resizebox{1.0\linewidth}{!}{
\begin{tabular}{cccccc}
\hline
\multirow{2}{*}{\begin{tabular}[c]{@{}c@{}}Mode\\ (Feature Num)\end{tabular}} & \multicolumn{4}{c}{Time} \\
        & \textbf{3h} & \textbf{6h} & \textbf{9h} & \textbf{24h} \\ \hline
Structure (5)            &   68.71  &  73.62  &   82.21   &  84.05    \\
Interaction (9)           &   73.62  &  83.44  &   84.66    &  84.05     \\
Text (11)                 &   73.62   &    87.73    &   84.05    &  85.28    \\ \cline{2-6} 
Phychology (13)        &   \begin{tabular}[c]{@{}c@{}}73.62{\footnotesize $\bm{+11.66}$}\\(\textbf{85.28})\end{tabular}   &    \begin{tabular}[c]{@{}c@{}}87.73{\footnotesize $\bm{+5.52}$}\\(\textbf{93.25})\end{tabular}    &   \begin{tabular}[c]{@{}c@{}}84.66{\footnotesize $\bm{+9.21}$}\\(\textbf{93.87})\end{tabular}   &  \begin{tabular}[c]{@{}c@{}}85.28{\footnotesize $\bm{+8.59}$}\\(\textbf{93.87})\end{tabular}     \\ \hline
\end{tabular}
}
\end{table}

The important findings described above also provide additional theoretical support for our realization of earlier conflict discovery. Because each comment of the data carries a posting time, we analyze the posts for different time periods. To do so, we sort all comments of each post by time, use the first 3 hours of comments for the computation of features, and evaluate their controversy detection performance. In the time interval, each round linearly increases the number of comments by 3 hours for multiple rounds of experiments, and in the features, each round adds different types of features on top of the existing ones, and the results are shown in Table \ref{tab: application}. It can be seen that around 3 hours after the post is sent out, the accuracy of detecting whether it is a controversial article or not can reach 85.28\% with the possession of psychological features, which is close to reaching the late state of other feature modes. Using only 6 hours of data approximated the performance of using 24 hours of data to achieve 93.25\% controversy detection performance. In addition to this, the detection performance of the psychological mode has a clear advantage in almost all stages, and the improvement becomes more obvious the further back we go (after "$+$" in the table are the values of the current column's improvement relative to the best accuracy). Our feature model generally achieves its best detection performance in the first and middle phases and remains stable until the event is unattended, reflecting the model's ability to recognize conflicts earlier as well as the stability of its features.

\section{Conclusion}
\label{sec: conclusion}
In this study, we collected data from various media platforms in both English and Chinese languages. Our main objective was to achieve high performance with minimal information by conducting experiments such as "one page" and early detection. By quantitatively evaluating the ability of each feature to discriminate conflicting and non-conflicting posts in LR, RF, and XGBoost classification algorithms, we try to find generalized and easily accessible features for controversy detection. We find that in comments, the difference in the number of likes between superior and inferior comments is an efficient and general feature to distinguish controversy. In conflicting posts, there is a high proportion of ascending gradients, i.e., replies get more likes than comments. Conversely, non-conflicting posts exhibit a lower proportion of ascending gradients. We also consider the difference in the number of replies per comment and then form two novel psychological features. These features achieve a general algorithmic enhancement. We discussed the emergence of the ascending gradient mechanism and its association with the backfire effect in ideological conflicts that have recently garnered attention. In terms of applications, we demonstrated that this difference in ascending gradient can be detected even by analyzing only a few hot comments. Meanwhile, our features enable controversy detection using minimal information without relying on complex semantic analysis or global social networks.

A limitation of this study is that all considered features are limited to local information, and we can directly find these features from the page of the post content. Although we show that among almost all local features, the psychological features are the most powerful for discriminating conflict, we do not further compare the AG and TAG with other global features (e.g., social network structure). At the same time, we also do not provide a direct causal correlation between the ascending gradient and conflict. The gap would be an exciting theoretical question: can psychological mechanisms help us find some local features that are more powerful than global features? Or must we rely on global datasets and complex computing models when solving computational sociology problems? In the future, we will take the conflict detection problem as an example to further explore how to solve computing problems in online media more quickly and effectively with limited data and computing power.


\appendix


\section{Features comparison}
\begin{figure*}[ht]
\centering
  \includegraphics[scale=0.6]{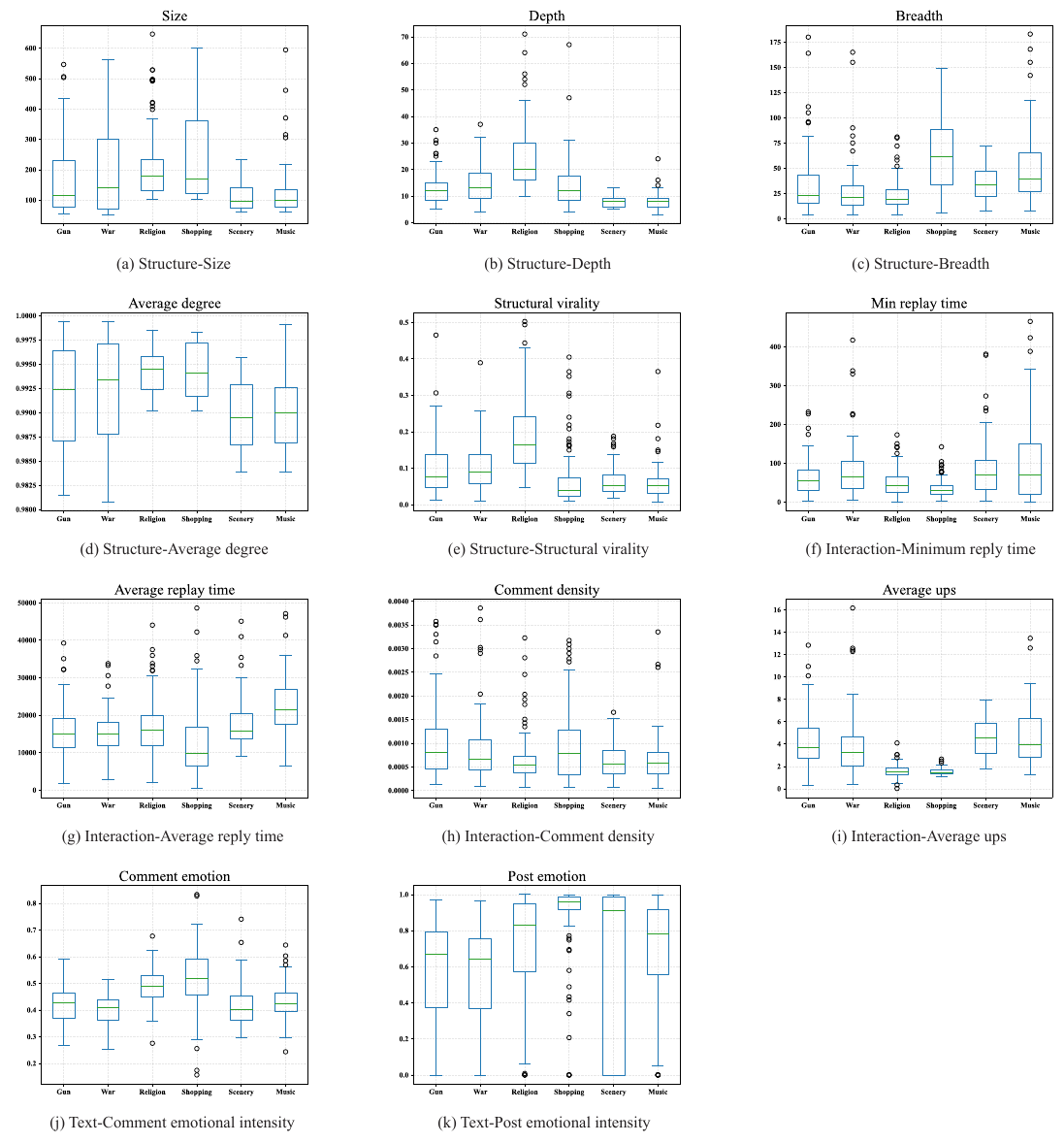}
  \caption{The box plot of each topic in 11 features (except ascending gradient and tier ascending gradient). Compare the differences in the distribution of controversial and non-controversial topics under the same features.}
  \label{fig: feature_compare}
\end{figure*}

We provide distributions of all features in Topic levels at Fig \ref{fig: feature_compare} as an appendix.
\clearpage

\section{Features description}
\begin{table}[ht]
\renewcommand\arraystretch{1.5}
\caption{Feature Description. }
\label{tab: features}

\resizebox{1.0\linewidth}{!}{
\begin{tabular}{ccccc}
    \toprule
    \textbf{No.} & \textbf{Name} & \textbf{Category} &\textbf{Symbol} & \textbf{Detailed description}\\
    \midrule
    1 & Size                & \textit{Structure} & $n$ & \makecell{Number of nodes in the comment tree} \\
    2 & Depth               & \textit{Structure} & $d$ & \makecell{Maximum depth of the comment tree} \\
    3 & Breadth             & \textit{Structure} & $b$ & \makecell{Maximum breadth of comment tree}    \\
    4 & Average degree      & \textit{Structure} & $k$ & \makecell{The average degree of the comment tree}  \\
    5 & Structural virality & \textit{Structure} & $v$ &  \makecell{The average shortest path of the comment tree}\\
    6 & Minimum reply time  & \textit{Interaction} & $t_{min}$ & \makecell{The shortest time of all reply pairs} \\
    7 & Average reply time  & \textit{Interaction} & $t_{avg}$ & \makecell{Average time for all reply pairs} \\
    8 & Comment density     & \textit{Interaction} & $c$ & \makecell{Number of comments per second on post}  \\
    9 & Average ups     & \textit{Interaction} & $q$ & \makecell{Average of the number of likes of all comments} \\
    10 & Comment emotional intensity & \textit{Text} & $s_c$ & \makecell{Average of the emotional intensity of all comments} \\
    11 & Post emotional intensity      & \textit{Text} & $s_p$ & \makecell{The emotional intensity of post content}  \\
    12 & Ascending gradient  & \textit{Psychology} & $p_a$ & \makecell{Percentage of difference in likes for all reply pairs} \\
    13 & Tier ascending gradient  & \textit{Psychology} & $p_t$ & \makecell{The difference in the number of reply to all comments}  \\
 	\bottomrule
\end{tabular}
}
\end{table}



\bibliographystyle{elsarticle-num} 
\bibliography{refs}





\end{document}